\documentclass[
aps,
prl,
twocolumn,
superscriptaddress,
nofootinbib
]{revtex4-2}

\usepackage{amsmath,amssymb,amsfonts,mathrsfs}
\usepackage{bm}
\usepackage{graphicx}
\usepackage{subfigure}
\usepackage{hyperref}
\usepackage{xcolor}

\begin{document}

\title{Completing the Penrose Process without a Horizon}

\author{Yang Huang}
\email{sps\_huangy@ujn.edu.cn}
\affiliation{
School of Physics and Technology, University of Jinan,
336 West Road of Nan Xinzhuang, Jinan 250022, Shandong, China
}

\author{Shiyang Hu}
\email{husy\_arcturus@163.com}
\affiliation{
School of Mathematics and Physics, University of South China,
Hengyang 421001, China
}

\author{Bin Chen}
\email{chenbin1@nbu.edu.cn}
\affiliation{
Institute of Fundamental Physics and Quantum Technology
and School of Physical Science and Technology,
Ningbo University, Ningbo, Zhejiang 315211, China
}
\affiliation{
School of Physics and Center for High Energy Physics,
Peking University, Beijing 100871, China
}

\author{Minyong Guo}
\email{minyongguo@bnu.edu.cn (Corresponding author)}
\affiliation{
School of Physics and Astronomy, Beijing Normal University,
Beijing 100875, China
}
\affiliation{
Key Laboratory of Multiscale Spin Physics
(Beijing Normal University), Ministry of Education,
Beijing 100875, China
}

\begin{abstract}
In its standard black-hole realization, the Penrose process uses an
event horizon to remove a negative-energy fragment. We show that, at
the kinematic level, horizon absorption can instead be replaced by
confinement within a compact ergoregion, without imposing an
absorbing or reflecting inner boundary. For a broad class of regular,
stationary, axisymmetric, asymptotically flat horizonless spacetimes,
we prove that every smooth connected component of the spatial
ergosurface is a compact torus, independently of the field equations
and matter content. Future-directed geodesics with negative Killing
energy encounter a forbidden neighborhood of each such boundary
component and are therefore confined. We derive an exterior
no-barrier condition and identify an open set of on-shell,
future-directed, four-momentum-conserving splittings producing both
a confined negative-energy fragment and an amplified partner. Under
equatorial reflection symmetry, once the latter enters the exterior
channel on an outward branch, it necessarily reaches infinity. A
rotating boson star explicitly realizes the complete process without
a horizon.
\end{abstract}

\maketitle

\textit{Introduction.---}
In the standard Penrose process, a particle decays inside the
ergoregion of a rotating black hole. One fragment crosses the event
horizon with negative Killing energy, allowing the other to escape
to infinity with more energy than the incident particle
\cite{Penrose:1969pc,Penrose:1971uk,Wald:1974kya}. The horizon thus
appears to play an essential role: it removes the negative-energy
product from the exterior. What, if anything, can fulfill this role
for a regular rotating object without a horizon?

The Penrose mechanism has since been extended to collisional,
electromagnetic, radiative, and repetitive realizations
\cite{Schnittman:2018ccg,Stuchlik:2021Penrose,
Ruffini:2024dwq}. Here we focus on a neutral two-body decay followed
by geodesic motion and ask whether it can be completed globally
without a horizon. An ergoregion admits future-directed
negative-energy states, but their existence alone does not complete
such a process. The negative-energy product must remain inaccessible
to infinity, the amplified partner must avoid exterior turning
barriers, and both outcomes must arise from the same on-shell,
future-directed, four-momentum-conserving decay. The Penrose process
in horizonless geometries has been related to superradiance and
ergoregion instabilities \cite{Vicente:2018mxl}. Horizonless
ergoregions have also been studied both in general settings
\cite{Friedman:1978ygc,Moschidis:2016zjy} and in specific models of
ultracompact stars, gravastars, and exotic compact objects
\cite{Cardoso:2007az,Chirenti:2008pf,Maggio:2017ivp}. These results
do not by themselves establish that one local particle splitting
simultaneously produces a globally confined negative-energy fragment
and an escaping amplified partner. Moreover, our confinement
mechanism requires no imposed absorbing or reflecting inner
boundary; it follows from the geometry and causal structure of the
smooth compact ergosurface.

Here we close this local-to-global gap under general geometric
conditions. We first prove that, for the class of regular,
stationary, axisymmetric, asymptotically flat horizonless spacetimes
considered here, every smooth connected component of the spatial
ergosurface is a compact torus. Although toroidal ergosurfaces occur
in explicit boson-star and related solutions
\cite{Kleihaus:2007vk,Herdeiro:2014jaa,Sun:2023ord,
Zhang:2023rwc,Huang:2024gtu,Liang:2026vls}, our classification is
independent of the field equations and matter content. We then show
that a future-directed causal fragment with conserved negative
Killing energy encounters a forbidden neighborhood of each such
boundary and cannot leave the corresponding compact ergoregion.

Finally, we derive an exterior no-barrier condition and prove that
it overlaps an open set of local Penrose splittings. The same
on-shell, future-directed, four-momentum-conserving decay can
therefore produce a confined negative-energy fragment and an
amplified partner in a mass-shell-allowed channel connected to
infinity. Under equatorial reflection symmetry, once the latter
enters this channel on an outward branch, it necessarily reaches the
asymptotic end. We realize all conditions explicitly in a rotating
boson-star spacetime.

\textit{Toroidal topology of horizonless ergosurfaces.---}
Consider a smooth, stationary, axisymmetric, asymptotically flat
spacetime without event horizons. Let
$\xi^\mu=(\partial_t)^\mu$ and
$\eta^\mu=(\partial_\phi)^\mu$ be the commuting stationary and axial
Killing fields. Assuming orthogonal transitivity, the metric can be
written as
\begin{equation}
\textrm{d}s^2
=
g_{tt}\textrm{d}t^2+2g_{t\phi}\textrm{d}t\,\textrm{d}\phi+g_{\phi\phi}\textrm{d}\phi^2
+g_{rr}\textrm{d}r^2+g_{\theta\theta}\textrm{d}\theta^2 .
\label{metric}
\end{equation}
The metric functions depend only on $(r,\theta)$. Away from the
regular rotation axis, we assume $g_{\phi\phi}>0$ and
\begin{equation}
D\equiv g_{t\phi}^{\,2}-g_{tt}g_{\phi\phi}>0,
\label{Killing-determinant}
\end{equation}
so that the $(t,\phi)$ Killing block is Lorentzian and
nondegenerate.

Let $\mathcal S$ be an asymptotically Euclidean spacelike slice, and
define the spatial ergoregion and ergosurface by
$\mathcal E=\{g_{tt}>0\}\subset\mathcal S$ and
$\Sigma=\partial\mathcal E$. Consider a smooth connected component
$\Sigma_i\subset\Sigma$, with
$\nabla_{\mathcal S}g_{tt}\neq0$ on $\Sigma_i$. Since
$g_{tt}\to-1$ at spatial infinity, $\mathcal E$ is contained in a
compact spatial domain; hence $\Sigma_i$ is compact.

It remains to determine whether $\Sigma_i$ can intersect the
rotation axis $\mathcal A=\{\eta^2=0\}$. Let $\rho$ be the proper
distance from a regular axis. Elementary flatness and smoothness give
\begin{equation}
g_{\phi\phi}=\rho^2+O(\rho^4),\,
g_{t\phi}=O(\rho^2),\,
D=-g_{tt}\big|_{\mathcal A}\rho^2+O(\rho^4).
\label{axis-expansion}
\end{equation}
Since the axis is regular and nondegenerate, we have 
$\lim_{\rho\to0}D/\rho^2>0$, which, together with the above expansion, implies
$g_{tt}|_{\mathcal A}<0$. Thus the stationary Killing field is timelike
on the regular axis, so $\Sigma_i\cap\mathcal A=\varnothing$.

Consequently, the axial $U(1)$ action is  free on $\Sigma_i$. The corresponding 
orbit space is a compact connected one-dimensional manifold without
boundary, hence is  $S^1$. Therefore, $\Sigma_i$ is a principal
$S^1$ bundle over $S^1$. Since all such bundles are topologically
trivial,
\begin{equation}
\Sigma_i\simeq S^1\times S^1=T^2 .
\label{torus}
\end{equation}
Thus every smooth connected component of the spatial ergosurface is
a compact, axis-free torus, independently of the field equations and
matter content. This supplies the closed boundary required for the
confinement argument below. Further details are given in
Sec.~\ref{sec:S-topology} of the Supplemental Material.

\textit{Local Penrose splitting.---}
We establish a local criterion for the simultaneous production of a
negative-energy fragment and an amplified positive-energy partner in
a single two-body decay. Let an incident particle with four-momentum
$p_0^\mu=m_0u_0^\mu$ decay at a point $x_*$ inside the ergoregion.
In its local rest frame, the fragment momenta can be written as
\begin{align}
p_1^\mu&=\varepsilon_1u_0^\mu+k n^\mu,\nonumber\\
p_2^\mu&=\varepsilon_2u_0^\mu-k n^\mu,
\label{local-decay}
\end{align}
where $n^\mu$ is a unit spacelike emission direction orthogonal to
$u_0^\mu$. Exact four-momentum conservation and the mass-shell
conditions $p_i^2=-m_i^2$ fix
\begin{align}
\varepsilon_1
&=\frac{m_0^2+m_1^2-m_2^2}{2m_0},\,
\varepsilon_2
=\frac{m_0^2+m_2^2-m_1^2}{2m_0},\nonumber\\
k
&=\frac{\sqrt{\lambda(m_0^2,m_1^2,m_2^2)}}{2m_0},
\label{two-body-kinematics}
\end{align}
where
$\lambda(a,b,c)=a^2+b^2+c^2-2ab-2ac-2bc$.
For $m_1+m_2<m_0$, one has $k>0$ and
$\varepsilon_i>0$. The fragment momenta are therefore on shell and
future directed, and satisfy
$p_0^\mu=p_1^\mu+p_2^\mu$ by construction.

To determine their Killing charges, decompose the stationary Killing
field in the incident-particle rest frame as
\begin{equation}
\xi^\mu=A u_0^\mu+\zeta^\mu,\,
A\equiv-\xi\cdot u_0=\frac{E_0}{m_0}>0,\,
\zeta\cdot u_0=0 .
\label{Killing-decomposition}
\end{equation}
Writing $s^2\equiv\zeta^2$, one has
$s^2=A^2+g_{tt}>A^2$ inside the ergoregion. The fragment energies are
\begin{align}
E_1(n)&=A\varepsilon_1-k\,\zeta\cdot n,\nonumber\\
E_2(n)&=A\varepsilon_2+k\,\zeta\cdot n
=E_0-E_1(n),
\label{fragment-energies}
\end{align}
while their angular momenta are
\begin{align}
L_1(n)&=\varepsilon_1\ell_0+k\,\eta\cdot n,\nonumber\\
L_2(n)&=\varepsilon_2\ell_0-k\,\eta\cdot n,\,
\ell_0\equiv\eta\cdot u_0=\frac{L_0}{m_0}.
\label{fragment-angular-momenta}
\end{align}
These charges obey
$E_1+E_2=E_0$ and $L_1+L_2=L_0$.

Fragment 1 has negative Killing energy when
\begin{equation}
k\,\zeta\cdot n>A\varepsilon_1 .
\label{negative-decay-directions}
\end{equation}
Since $\max_n(\zeta\cdot n)=s$ over the unit emission sphere, such
directions exist if and only if
\begin{equation}
ks>A\varepsilon_1 .
\label{negative-decay-existence}
\end{equation}
Whenever this inequality is strict, the allowed directions form an
open set and satisfy
\begin{equation}
E_1<0,\, E_2=E_0-E_1>E_0 .
\label{local-energy-amplification}
\end{equation}
Hence a single on-shell, future-directed,
four-momentum-conserving decay can produce a negative-energy
fragment and an amplified partner. We next determine the global fate
of these same decay products.

\textit{Negative-energy confinement.---}
To track the fragments beyond the decay event, define
$\omega=-g_{t\phi}/g_{\phi\phi}$ and
$\alpha^2=D/g_{\phi\phi}$. In the regular orbit space, the mass-shell
condition takes the form
\begin{equation}
\mathcal V(E,L;m)
\equiv
\frac{(E-\omega L)^2}{\alpha^2}
-m^2-\frac{L^2}{g_{\phi\phi}}
=
g_{rr}(p^r)^2+g_{\theta\theta}(p^\theta)^2 .
\label{mass-shell-function}
\end{equation}
Physical motion requires $\mathcal V\geq0$; null geodesics  are recovered
by setting $m=0$. Furthermore since 
$p^t=(E-\omega L)/\alpha^2$, future-directedness imposes
$E-\omega L>0$.

For the on-shell fragment momenta,
$\mathcal V_i(x_*)\geq0$, so both products define admissible
geodesic initial data. Strict positivity selects the nonturning
subset; in the explicit example below, fragment 1 is instead
produced at its outer radial turning point.

Now consider fragment 1 with $E_1<0$. On the ergosurface,
$g_{tt}=0$ implies $\alpha^2=\omega^2g_{\phi\phi}$, while
future-directedness gives $\omega L_1<E_1<0$. Hence
\begin{equation}
\mathcal V_1\big|_{\Sigma}
=
\frac{E_1^2-2\omega E_1L_1}{\alpha^2}-m_1^2
<
-\frac{E_1^2}{\alpha^2}-m_1^2<0 .
\label{negative-barrier}
\end{equation}
The same conclusion holds for $m_1=0$. Since physical motion
requires $\mathcal V_1\geq0$, the fragment cannot reach a smooth
ergosurface component. Compactness and continuity extend the strict
inequality in Eq.~\eqref{negative-barrier} to a forbidden
neighborhood of the entire component. Applying the argument to every
boundary component, a fragment with conserved $E_1<0$ cannot leave
the corresponding compact ergoregion. Ergosurface confinement
therefore replaces horizon absorption at the kinematic level.

\textit{Escape of the amplified fragment.---}
Having established confinement for the negative-energy fragment, we now turn to the amplified partner. While local energy amplification is necessary, it is not sufficient: fragment 2 must additionally be able to propagate to infinity without encountering an exterior mass-shell barrier. The future-directed branch of
Eq.~\eqref{mass-shell-function} gives
\begin{equation}
E_2\geq
\omega L_2+\alpha
\sqrt{m_2^2+\frac{L_2^2}{g_{\phi\phi}}}.
\label{escape-branch}
\end{equation}
Let $\mathcal D$ be a connected exterior domain containing the
splitting point, bounded away from the rotation axis, and extending
to an asymptotically flat end. Regularity and asymptotic flatness
make
\begin{equation}
\alpha_{\max}\equiv\sup_{\mathcal D}\alpha,\,
C_{\max}\equiv\sup_{\mathcal D}
\left(|\omega|+\frac{\alpha}{\sqrt{g_{\phi\phi}}}\right)
\label{global-bounds}
\end{equation}
finite. Using
\[
\sqrt{m_2^2+\frac{L_2^2}{g_{\phi\phi}}}
\leq m_2+\frac{|L_2|}{\sqrt{g_{\phi\phi}}},
\]
we obtain the sufficient no-barrier condition
\begin{align}
E_2&>\alpha_{\max}m_2+C_{\max}|L_2|,\nonumber\\
|b_2|&<
\frac{1-\alpha_{\max}m_2/E_2}{C_{\max}},\,
b_2\equiv\frac{L_2}{E_2}.
\label{escape-condition}
\end{align}
The second line is the equivalent impact-parameter form. When
satisfied strictly, either form guarantees $\mathcal V_2>0$
throughout $\mathcal D$ and therefore defines an open
mass-shell-allowed channel connecting the splitting point to the
asymptotic end.

\textit{Confinement and amplified escape from the same splitting.---}
The central question is whether one emission direction can produce a
negative-energy fragment while placing its amplified partner in the
exterior no-barrier channel. To expose the directional freedom, define
\begin{equation}
e^\mu\equiv\frac{\zeta^\mu}{s},\,
n^\mu=\cos\chi\,e^\mu+\sin\chi\,q^\mu,
\label{angular-emission-direction}
\end{equation}
where $q\cdot u_0=0,\, q\cdot e=0,\, q^2=1$.
As $q^\mu$ ranges over the transverse unit circle,
Eq.~\eqref{angular-emission-direction} parametrizes the local
emission sphere. In particular,
\begin{align}
\zeta\cdot n
&=s\cos\chi,\nonumber\\
\eta\cdot n
&=
\frac{\eta\cdot\zeta}{s}\cos\chi
+(\eta\cdot q)\sin\chi .
\label{angular-projections}
\end{align}
The negative-energy and no-barrier requirements become
\begin{align}
ks\cos\chi
&>A\varepsilon_1,\nonumber\\
A\varepsilon_2+ks\cos\chi
&>
\alpha_{\max}m_2
+C_{\max}
\left|
\varepsilon_2\ell_0-k\,\eta\cdot n
\right|.
\label{angular-compatibility}
\end{align}
Because the inequalities are strict and continuous, compatible
directions form an open subset of the local emission sphere; see
Sec.~\ref{sec:S-openness} of the Supplemental Material for the
complete two-angle parametrization.

A directly testable sufficient criterion follows from
$\chi=0$, which minimizes $E_1$ at fixed incident momentum and
fragment masses. Equation~\eqref{angular-compatibility}
then reduces, for finite fragment masses, to
\begin{align}
ks&>A\varepsilon_1,\nonumber\\
C_{\max}
\left|s\varepsilon_2\ell_0-k\,\eta\cdot\zeta\right|
&<
s\left(A\varepsilon_2+ks-\alpha_{\max}m_2\right).
\label{massive-compatibility}
\end{align}
Whenever these strict inequalities hold, the same on-shell,
future-directed, four-momentum-conserving decay produces a confined
negative-energy fragment and an amplified partner in the no-barrier
channel.

The small-fragment-mass limit makes the geometric content
transparent. As $m_1,m_2\to0^+$,
$\varepsilon_1,\varepsilon_2,k\to m_0/2$. The first inequality in
Eq.~\eqref{massive-compatibility} reduces to $s>A$, which is
automatic inside the ergoregion, while the second becomes
\begin{equation}
C_{\max}
\left|s\ell_0-\eta\cdot\zeta\right|
<s(A+s).
\label{geometric-compatibility}
\end{equation}
Thus Eq.~\eqref{geometric-compatibility} is a simple sufficient
condition for the local negative-energy cone to overlap the exterior
no-barrier window. If it holds strictly, continuity gives a nonempty
open set of sufficiently small positive fragment masses and nearby
emission directions for which
\begin{equation}
E_1<0,\, E_2=E_0-E_1>E_0,\,
\mathcal V_2>0
\label{central-compatibility}
\end{equation}
throughout $\mathcal D$.
For generic two-dimensional motion, a no-barrier channel need not
guarantee escape of a specified orbit. With equatorial reflection
symmetry, however, a fragment emitted tangent to the equatorial
plane remains there. Once it enters $\mathcal D$ on an outward
branch, $g_{rr}(p_2^r)^2=\mathcal V_2>0$ prevents any further radial
turning point. In the explicit trajectory below, the fragment is
initially inward, turns at an inner radius, and then follows this
outward channel to the asymptotic end.

\begin{figure*}[!t]
\centering
\subfigure[]{
\includegraphics[
width=0.45\textwidth
]{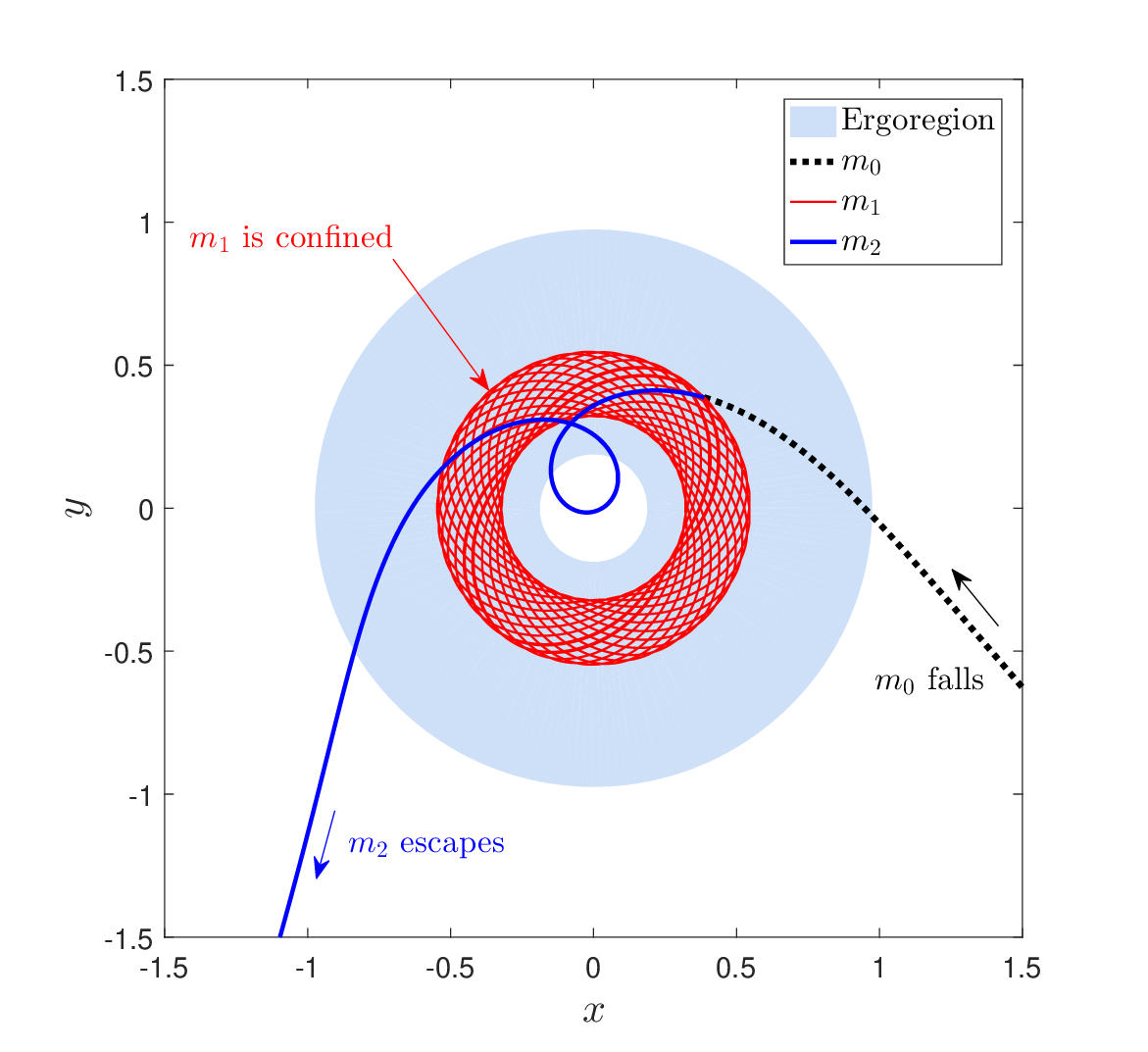}
}
\hfill
\subfigure[]{
\includegraphics[
width=0.45\textwidth
]{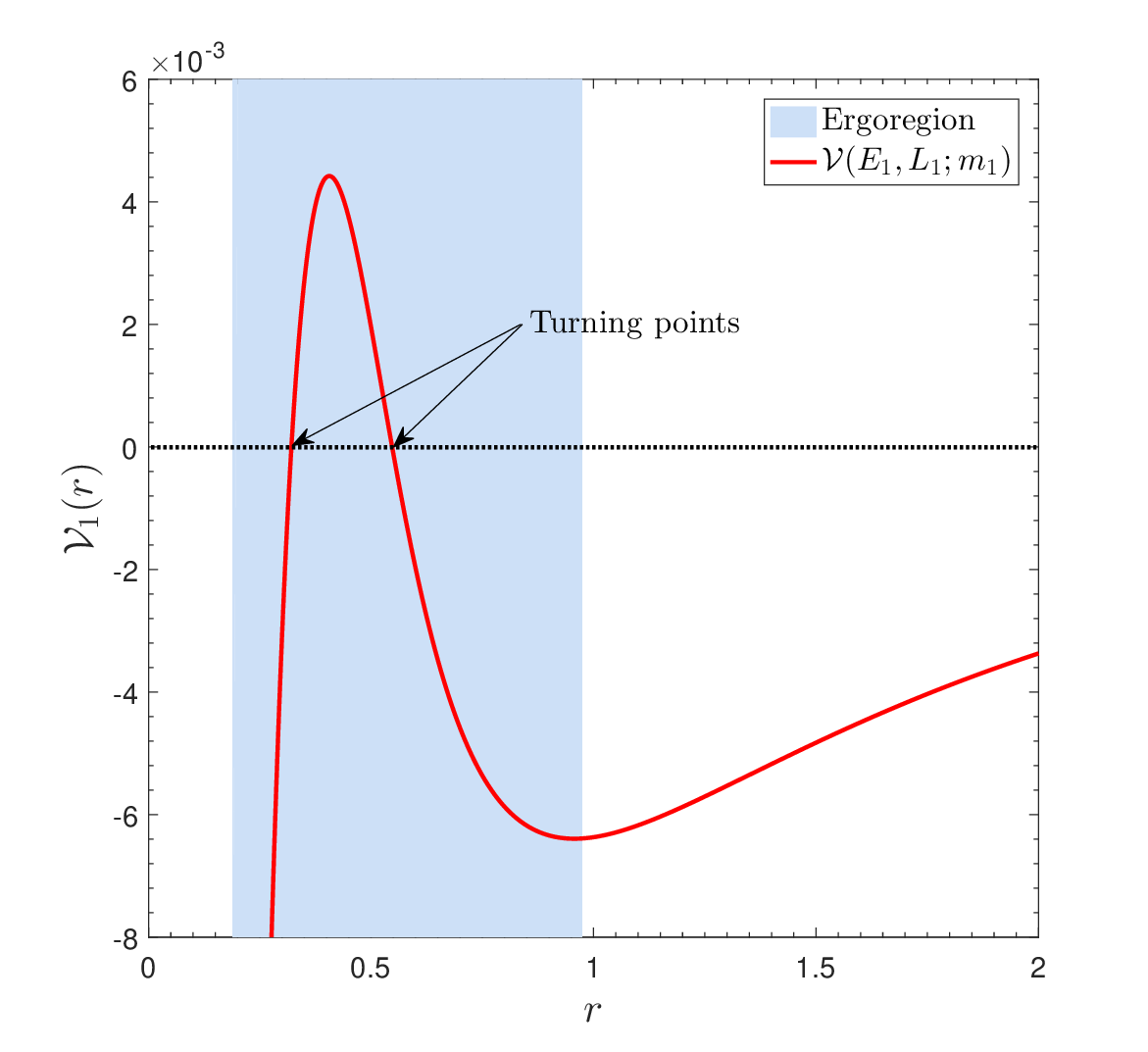}
}
\caption{
\textit{Horizonless Penrose process from one local decay.}
(a) An incident particle reaches a splitting point $x_*$ inside the
toroidal ergoregion and decays into a negative-energy fragment,
$E_1<0$, and an amplified partner,
$E_2=E_0-E_1>E_0$. The negative-energy fragment remains confined.
The amplified fragment is initially directed inward, reaches an
inner radial turning point, and subsequently enters the exterior
no-barrier channel on an outward branch, reaching infinity. The
decay satisfies exact four-momentum conservation, the mass-shell
conditions, and future-directedness.
(b) Mass-shell function $\mathcal V_1$ of the negative-energy
fragment. Its allowed interval is bounded entirely inside the
ergoregion.
}
\label{fig:boson-star}
\end{figure*}

\textit{Rotating-boson-star realization.---}
We demonstrate the mechanism in a regular rotating boson star with a
compact toroidal ergoregion and equatorial reflection symmetry. The background construction and complete splitting data are given in
Secs.~\ref{sec:S-background} and \ref{sec:S-splitting} of the
Supplemental Material.

An incident particle with $(m_0,E_0,L_0)=(1,1,0)$ reaches an
equatorial splitting point inside the ergoregion and decays into
fragments with $m_1=0.01$ and $m_2=0.0757599$. Their Killing charges
are
\begin{equation}
(E_1,L_1)=(-0.0214,-0.194),\,
(E_2,L_2)=(1.0214,0.194).
\label{example-charges}
\end{equation}
The full numerical momenta satisfy
$p_0^\mu=p_1^\mu+p_2^\mu$, $p_i^2=-m_i^2$,
$p_i^t>0$, and $\mathcal V_i(x_*)\geq0$, and therefore define one
on-shell, future-directed, four-momentum-conserving, locally
admissible decay.

For numerical transparency, we choose the emission direction so that
$p_1^r(x_*)=0$, placing fragment 1 at its outer radial turning point,
$\mathcal V_1(x_*)=0$. This turning-point representative makes the
confined allowed interval directly visible but is not required by
the mechanism. The same numerical direction satisfies the two
compatibility conditions with the strict margins
\begin{align}
\delta_-
&\equiv
k\,\zeta\cdot n-A\varepsilon_1
=-E_1
\simeq0.0214>0,
\nonumber\\
\delta_{\rm esc}
&\equiv
E_2-\alpha_{\max}m_2-C_{\max}|L_2|
\simeq0.874>0 .
\label{example-compatibility-margins}
\end{align}
The negative-energy and no-barrier conditions therefore persist for
nearby emission directions. The angular characterization of the
chosen direction and the corresponding openness argument are given
in Secs.~\ref{sec:S-example-direction} and
\ref{sec:S-openness} of the Supplemental Material.

Equation~\eqref{negative-barrier} and the allowed interval shown in
Fig.~\ref{fig:boson-star}(b) confine fragment 1 within the
ergoregion. Fragment 2 is initially directed inward, reaches an
inner radial turning point, and then enters the exterior no-barrier
channel on its outward branch. No further radial turning point
occurs, and it reaches infinity with
\begin{equation}
\Delta E=E_2-E_0=-E_1=0.0214,\,
\eta=\frac{\Delta E}{E_0}\simeq2.14\%.
\label{example-efficiency}
\end{equation}
This explicitly realizes negative-energy confinement and amplified
escape from the same local splitting.

\textit{Conclusions.---}
We have completed the Penrose process without a horizon at the
kinematic level. Under the geometric assumptions stated above, every
smooth connected component of the spatial ergosurface is a compact,
axis-free torus. This classification is independent of the field
equations and matter content. More importantly, future-directed
causal geodesics with conserved negative Killing energy encounter a
forbidden neighborhood of each such boundary component and therefore
cannot leave the corresponding compact ergoregion. The compact
ergoregion can therefore play the kinematic role normally assigned
to a horizon: rather than absorbing the negative-energy product, it
keeps that product inaccessible to the asymptotic region. No
absorbing or reflecting inner boundary is required; confinement
follows from the geometry and causal structure of the smooth compact
ergosurface.

We further derived a sufficient exterior no-barrier condition and
proved that it is compatible with a nonempty open set of on-shell,
future-directed, four-momentum-conserving Penrose splittings. The
central result is that negative-energy confinement and amplified
escape need not be arranged through independently selected
trajectories; rather, they arise from the two products of the same
local decay. For generic two-dimensional motion, the no-barrier
condition identifies a mass-shell-allowed channel connected to
infinity. Under equatorial reflection symmetry, once the amplified
fragment enters this channel on an outward branch, strict positivity
of its mass-shell function rules out any subsequent radial turning
point, and the fragment necessarily reaches infinity. The
rotating-boson-star example realizes this complete chain explicitly:
one fragment carries negative Killing energy and remains confined,
whereas its partner escapes with more Killing energy than the
incident particle.

Our result establishes test-particle Killing-energy amplification,
but does not establish the self-consistent dynamical extraction of
energy and angular momentum from the central object. In particular,
it does not determine how the confined fragment subsequently
interacts with the bosonic matter, whether its negative Killing
energy is absorbed or redistributed, or how backreaction changes the
mass, angular momentum, and ergoregion of the spacetime. Repeated
splittings may deform or eliminate the ergoregion and thereby modify
the confinement and escape channels themselves. A self-consistent
evolution of the matter and geometry is required to address these
questions and to determine their connection to ergoregion
instabilities in horizonless compact objects
\cite{Friedman:1978ygc,Cardoso:2007az,
Chirenti:2008pf,Maggio:2017ivp}
and to the subsequent nonlinear dynamics
\cite{Siemonsen:2025wib,Siemonsen:2025fne}.

\begin{acknowledgments}
We thank Guo-Ping Li and Ke-Jian He for their hospitality and
Yong-Qiang Wang and Yehui Hou for helpful discussions. This work was
supported by NSFC under Grant Nos.~12575048, 12275004, and 12588101,
and by the Shandong Provincial Natural Science Foundation under Grant
No.~ZR2024QA032. M.G. also acknowledges support from the BNU Tang
Scholar Program.
\end{acknowledgments}

\bibliography{references}


\clearpage
\onecolumngrid

\begin{center}
\textbf{\large Supplemental Material for}\\[1mm]
\textbf{\large
``Completing the Penrose Process without a Horizon''
}
\end{center}

\setcounter{secnumdepth}{1}

\setcounter{section}{0}
\setcounter{equation}{0}
\setcounter{figure}{0}
\setcounter{table}{0}

\renewcommand{\thesection}{\Roman{section}}
\renewcommand{\theequation}{S\arabic{equation}}
\renewcommand{\thefigure}{S\arabic{figure}}
\renewcommand{\thetable}{S\arabic{table}}

\section{Geometric assumptions and ergosurface topology}
\label{sec:S-topology}

Let $\mathcal S$ be an asymptotically Euclidean spacelike slice and
define
$\mathcal E=\{g_{tt}>0\}\subset\mathcal S$ and
$\Sigma=\partial\mathcal E$. Consider a connected component
$\Sigma_i\subset\Sigma$ on which zero is a regular value of
$g_{tt}|_{\mathcal S}$:
\begin{equation}
g_{tt}\big|_{\Sigma_i}=0,\,
\nabla_{\mathcal S}g_{tt}\big|_{\Sigma_i}\neq0 .
\label{S-regular-ergosurface}
\end{equation}
The regular-value theorem ensures that $\Sigma_i$ is a smooth
embedded two-manifold without boundary.

Asymptotic flatness gives $g_{tt}\to-1$ at spatial infinity.
Therefore there exists a compact set $K\subset\mathcal S$ outside
which $g_{tt}<0$, and hence $\mathcal E\subset K$. Since each
connected component of the closed set $\Sigma=\partial\mathcal E$
is closed, $\Sigma_i$ is a closed subset of $K$ and is therefore
compact.

Let $\mathcal A=\{\eta^2=0\}$ be the rotation axis and let $\rho$ be
the proper distance from a regular portion of $\mathcal A$.
Elementary flatness and smoothness give
\begin{equation}
g_{\phi\phi}=\rho^2+O(\rho^4),\,
g_{t\phi}=O(\rho^2).
\label{S-axis-expansions}
\end{equation}
Thus
\begin{equation}
D=g_{t\phi}^{\,2}-g_{tt}g_{\phi\phi}
=-g_{tt}\big|_{\mathcal A}\rho^2+O(\rho^4).
\label{S-D-expansion}
\end{equation}
The regular-axis nondegeneracy condition
$\lim_{\rho\to0}D/\rho^2>0$ implies
$g_{tt}|_{\mathcal A}<0$, and hence
\begin{equation}
\Sigma_i\cap\mathcal A=\varnothing .
\label{S-axis-disjoint}
\end{equation}

The induced $U(1)$ action on $\Sigma_i$ is therefore free. Its
quotient $B_i=\Sigma_i/U(1)$ is a smooth, compact, connected
one-dimensional manifold. Since the action is locally a product and
$\Sigma_i$ has no boundary, neither does $B_i$. Hence
$B_i\simeq S^1$.

The projection $\Sigma_i\to B_i$ defines a principal $U(1)$ bundle.
Such bundles are classified by their first Chern class in
$H^2(B_i,\mathbb Z)$. Since $H^2(S^1,\mathbb Z)=0$, the bundle is
trivial:
\begin{equation}
\Sigma_i\simeq S^1\times S^1=T^2 .
\label{S-torus}
\end{equation}
The conclusion applies componentwise; an ergoregion with several
smooth boundary components has a boundary given by a disjoint union
of tori.

The regular-value condition is essential. If
$\nabla_{\mathcal S}g_{tt}=0$ at a zero of $g_{tt}$, the level set
need not be a smooth surface: components may touch, merge, or undergo
a topology-changing pinch-off. Such critical configurations are not
covered by the classification above.

\section{General emission directions and openness}
\label{sec:S-openness}
We complete the angular parametrization introduced in
Eq.~\eqref{angular-emission-direction} and make the openness of the
compatible emission directions explicit. Choose an orthonormal
spatial triad $\{e^\mu,q_1^\mu,q_2^\mu\}$ in the
incident-particle rest space, with
\begin{equation}
e^\mu=\frac{\zeta^\mu}{s},\,
q_A\cdot u_0=0,\,
q_A\cdot e=0,\,
q_A\cdot q_B=\delta_{AB}.
\label{S-rest-triad}
\end{equation}
Every unit emission direction can be written as
\begin{align}
n^\mu(\chi,\beta)
&=
\cos\chi\,e^\mu+\sin\chi\,q^\mu(\beta),\nonumber\\
q^\mu(\beta)
&=
\cos\beta\,q_1^\mu+\sin\beta\,q_2^\mu .
\label{S-general-emission-direction}
\end{align}
This parametrization automatically satisfies
$n^2=1$ and $n\cdot u_0=0$.

The relevant projections are
\begin{align}
\zeta\cdot n
&=s\cos\chi,\nonumber\\
\eta\cdot n
&=
\frac{\eta\cdot\zeta}{s}\cos\chi
+\eta\cdot q(\beta)\sin\chi .
\label{S-general-projections}
\end{align}
Define
\begin{align}
F_-(\chi,\beta)
&=
ks\cos\chi-A\varepsilon_1,\nonumber\\
F_{\rm esc}(\chi,\beta)
&=
A\varepsilon_2+ks\cos\chi-\alpha_{\max}m_2
-C_{\max}
\left|
\varepsilon_2\ell_0
-k\left[
\frac{\eta\cdot\zeta}{s}\cos\chi
+\eta\cdot q(\beta)\sin\chi
\right]
\right|.
\label{S-compatibility-functions}
\end{align}
Compatible emission directions satisfy
\begin{equation}
F_-(\chi,\beta)>0,\,
F_{\rm esc}(\chi,\beta)>0 .
\label{S-compatible-directions}
\end{equation}
Since these are strict inequalities of continuous functions, every
solution belongs to an open subset of the local emission sphere.

The direction $\chi=0$ maximizes $\zeta\cdot n$ and therefore
minimizes $E_1$ at fixed incident momentum and fragment masses:
\begin{equation}
E_{1,\min}=A\varepsilon_1-ks .
\label{S-minimum-energy}
\end{equation}
It is thus the locally optimal direction for negative-energy
production. It need not optimize the full constrained problem,
because changing $(\chi,\beta)$ also changes $L_2$ and the
meridional fragment momenta.

The local mass-shell function satisfies
\begin{equation}
\mathcal V_i(x_*)
=
g_{rr}(p_i^r)^2+g_{\theta\theta}(p_i^\theta)^2 .
\label{S-local-mass-shell}
\end{equation}
Strict positivity characterizes the nonturning subset and is stable
under small changes of the decay data. The explicit negative-energy
trajectory below is a turning-point representative with
$\mathcal V_1(x_*)=0$; the analytic openness result does not rely on
this special choice.

For the simplest guaranteed-escape subset, restrict to equatorial
emission with $p_2^r>0$. These strict conditions define an open
subset of the equatorial decay data. The explicit trajectory below
is instead initially inward; after reaching an inner radial turning
point, it enters the exterior channel on an outward branch.

\section{Rotating boson-star background}
\label{sec:S-background}

We consider Einstein gravity minimally coupled to a complex massive
scalar field,
\begin{equation}
S
=
\int \textrm{d}^4x\,\sqrt{-g}
\left[
\frac{R}{16\pi}
-\nabla_\mu\Psi^*\nabla^\mu\Psi
-\mu^2|\Psi|^2
\right].
\label{S-action}
\end{equation}
The rotating boson-star ansatz is
\begin{equation}
\textrm{d}s^2
=
-e^{f_0}\textrm{d}t^2
+e^{f_1}(\textrm{d}r^2+r^2\textrm{d}\theta^2)
+e^{f_2}r^2\sin^2\theta(\textrm{d}\phi-W\textrm{d}t)^2 ,
\label{S-boson-metric}
\end{equation}
with scalar field
\begin{equation}
\Psi=\psi(r,\theta)e^{i(m_s\phi-\omega_s t)} .
\label{S-scalar-ansatz}
\end{equation}
The functions $f_0,f_1,f_2,W$, and $\psi$ depend on $(r,\theta)$.

We use a regular $m_s=1$ solution with a compact toroidal
ergoregion, obtained numerically following
Refs.~\cite{Herdeiro:2015gia,Huang:2024gtu,Kleihaus:2007vk}.
Its parameters are summarized in Table~\ref{tab:S-background}.

\begin{table}[h]
\caption{Boson-star parameters used in the explicit realization.}
\begin{ruledtabular}
\begin{tabular}{lc}
Quantity & Value \\
\hline
scalar frequency $\omega_s$ & $0.65$ \\
ADM mass $M$ & $1.02$ \\
ADM angular momentum $J$ & $0.94$ \\
inner equatorial ergosurface radius & $0.19$ \\
outer equatorial ergosurface radius & $0.97$
\end{tabular}
\end{ruledtabular}
\label{tab:S-background}
\end{table}

For Eq.~\eqref{S-boson-metric},
\begin{align}
g_{tt}
&=-e^{f_0}+e^{f_2}r^2\sin^2\theta\,W^2,\nonumber\\
g_{t\phi}
&=-e^{f_2}r^2\sin^2\theta\,W,\nonumber\\
g_{\phi\phi}
&=e^{f_2}r^2\sin^2\theta .
\label{S-metric-components}
\end{align}
The ergosurface is determined by $g_{tt}=0$. For the solution used
here, it is a smooth compact torus with the equatorial intersections
listed in Table~\ref{tab:S-background}.

\section{Complete local splitting data}
\label{sec:S-splitting}

The splitting occurs on the equatorial plane at
\begin{equation}
x_*=(r_*,\pi/2),\, r_*=0.547324 .
\label{S-splitting-point}
\end{equation}
The nonvanishing metric components at $x_*$ are
\begin{align}
g_{tt}&=0.12417579,\, g_{rr}=9.10695644,\nonumber\\
g_{\theta\theta}&=2.72811712,\,
g_{\phi\phi}=3.01678778,\,
g_{t\phi}=-0.72627334 .
\label{S-metric-at-splitting}
\end{align}

The incident particle has
\begin{equation}
(m_0,E_0,L_0)=(1,1,0).
\label{S-incident-data}
\end{equation}
Numerical integration confirms that it reaches $x_*$ from the
asymptotically flat region. At the splitting point, its four-velocity
and the unit emission direction are
\begin{align}
u_0^\mu
&=
(19.735503566986445,-1.434320004222809,0,
4.751202629642263),\nonumber\\
n_*^\mu
&=
(-18.739294442182203,1.434610219434106,0,
-4.640478608488062).
\label{S-local-vectors}
\end{align}
We denote the numerical emission direction by $n_*^\mu$ to
distinguish it from a general direction on the local emission sphere.
The orthonormality residuals are
\begin{equation}
|u_0^2+1|=5.68\times10^{-14},\,
|u_0\cdot n_*|=4.26\times10^{-13},\,
|n_*^2-1|=3.13\times10^{-13}.
\label{S-orthonormal-residuals}
\end{equation}

For
\begin{equation}
m_1=0.01,\, m_2\simeq0.0757599,
\label{S-fragment-masses}
\end{equation}
the two-body coefficients are
\begin{equation}
\varepsilon_1\simeq0.49718,\,
\varepsilon_2\simeq0.50282,\,
k\simeq0.49708 .
\label{S-decay-coefficients}
\end{equation}
The fragment momenta constructed from
Eq.~\eqref{local-decay} are
\begin{align}
p_1^\mu
&=
(0.497180219768636,0,0,0.05551652011001657),
\nonumber\\
p_2^\mu
&=
(19.238323347227055,-1.4343200042234983,0,
4.695686109534503).
\label{S-fragment-momenta}
\end{align}

The largest four-momentum-conservation residual is
\begin{equation}
\max_\mu|p_0^\mu-p_1^\mu-p_2^\mu|
=9.25\times10^{-12},
\label{S-momentum-residual}
\end{equation}
while the mass-shell residuals are
\begin{equation}
|p_1^2+m_1^2|=8.61\times10^{-17},\,
|p_2^2+m_2^2|=8.60\times10^{-14}.
\label{S-mass-shell-residuals}
\end{equation}

The corresponding Killing charges are
\begin{equation}
(E_1,L_1)\simeq(-0.0214,-0.194),\,
(E_2,L_2)\simeq(1.0214,0.194).
\label{S-fragment-charges}
\end{equation}
Their conservation residuals are
\begin{equation}
|E_0-E_1-E_2|=4.91\times10^{-13},\,
|L_0-L_1-L_2|=9.30\times10^{-14}.
\label{S-charge-residuals}
\end{equation}

Future-directedness and local mass-shell admissibility are verified
by
\begin{align}
p_1^t&\simeq0.497>0,\,
p_2^t=19.238>0,\nonumber\\
\mathcal V_1(x_*)&=0,\nonumber\\
 \mathcal V_2(x_*)&\simeq18.736>0 .
\label{S-local-admissibility}
\end{align}
Here $\mathcal V_i$ is evaluated using the physical-momentum
convention of Eq.~\eqref{mass-shell-function}. Thus both fragments
define admissible geodesic initial data. Fragment 1 is produced at a
radial turning point, identified below as its outer turning point,
whereas fragment 2 is initially directed inward:
\begin{equation}
p_2^r(x_*)\simeq-1.434<0 .
\label{S-initial-radial-direction}
\end{equation}

\section{Geometric characterization of the chosen direction}
\label{sec:S-example-direction}

For the explicit splitting, $A=E_0/m_0=1$ and
\begin{equation}
\zeta^\mu=\xi^\mu-u_0^\mu,\,
s=\sqrt{\zeta^2}=\sqrt{1+g_{tt}(x_*)}.
\label{S-example-zeta}
\end{equation}
Defining $e^\mu=\zeta^\mu/s$, the equatorial emission direction can
be decomposed as
\begin{equation}
n_*^\mu
=
\cos\chi_*\,e^\mu+\sin\chi_*\,q_{\rm eq}^\mu,
\label{S-example-angular-direction}
\end{equation}
where
\begin{equation}
q_{\rm eq}\cdot u_0=0,\,
q_{\rm eq}\cdot e=0,\,
q_{\rm eq}^2=1,\,
q_{\rm eq}^\theta=0 .
\label{S-equatorial-transverse-vector}
\end{equation}
The angle and transverse direction are determined by
\begin{align}
\cos\chi_*
&=
\frac{\zeta\cdot n_*}{s}
\simeq0.984,\,
\chi_*\simeq0.057\pi,\nonumber\\
q_{\rm eq}^\mu
&=
\frac{n_*^\mu-\cos\chi_*\,e^\mu}{\sin\chi_*}
\simeq(-19.250, 1.474, 0, -2.923).
\label{S-example-angle}
\end{align}
Thus $n_*^\mu$ is an equatorial tilt away from
$e^\mu=\zeta^\mu/s$, the direction that locally minimizes $E_1$ at
fixed incident momentum and fragment masses.

For numerical transparency, the tilt is selected so that
\begin{equation}
p_1^r(x_*)
=
\varepsilon_1u_0^r+k n_*^r=0 .
\label{S-turning-direction}
\end{equation}
This places fragment 1 at a radial turning point and makes its
confined allowed interval directly visible. The potential analysis
below identifies this point as the outer turning point,
$r_{1,+}=r_*$. Radial momentum conservation then gives
\begin{equation}
p_2^r(x_*)=p_0^r(x_*)<0,
\label{S-radial-momentum-conservation}
\end{equation}
so fragment 2 is initially inward.

The turning-point condition in
Eq.~\eqref{S-turning-direction} is imposed only to provide a
transparent numerical representative; it is not required by the
general confinement-and-escape mechanism. To verify compatibility
for the same numerical direction, define the strict margins
\begin{align}
\delta_-
&\equiv
k\,\zeta\cdot n_*-A\varepsilon_1
=-E_1
=0.0214>0,\nonumber\\
\delta_{\rm esc}
&\equiv
E_2-\alpha_{\max}m_2-C_{\max}|L_2|
=0.874>0 .
\label{S-example-compatibility-margins}
\end{align}
The first margin verifies negative-energy production, while the
second places the amplified fragment in the exterior no-barrier
window. Their strict positivity shows that these two compatibility
conditions persist under sufficiently small changes of the emission
direction. Nearby compatible directions need not satisfy
$p_1^r(x_*)=0$ and generically have
$\mathcal V_1(x_*)>0$. The complete two-angle parametrization and
the corresponding openness argument are given in
Sec.~\ref{sec:S-openness}.

For reference, the unconstrained locally optimal direction
$\chi=0$ gives
\begin{equation}
E_{1,\min}=A\varepsilon_1-ks,\,
\Delta E_{\max}^{\rm loc}=ks-A\varepsilon_1 .
\label{S-local-maximum-gain}
\end{equation}
The chosen direction sacrifices part of this unconstrained local
gain to realize a particularly transparent turning-point geometry.
It is not asserted to maximize the extraction efficiency under all
global confinement and escape constraints.

\section{Confinement and escape checks}
\label{sec:S-checks}

For fragment 1, direct evaluation gives the equatorial turning points
\begin{equation}
r_{1,-}=0.320998,\, r_{1,+}=r_*=0.547324 .
\label{S-negative-turning-points}
\end{equation}
Both lie inside the ergoregion:
\begin{equation}
r_{\rm ergo}^{\rm inner}
<
r_{1,-}
<
r_{1,+}
<
r_{\rm ergo}^{\rm outer}.
\label{S-negative-contained}
\end{equation}
The allowed radial interval is therefore bounded entirely within the
toroidal ergoregion, in agreement with the general barrier
Eq.~\eqref{negative-barrier}.

For fragment 2, the exterior domain gives
\begin{equation}
\alpha_{\max}=1,\, C_{\max}=0.370343 .
\label{S-exterior-bounds}
\end{equation}
Its impact parameter and the sufficient no-barrier bound are
\begin{equation}
|b_2|=0.189548,\,
\frac{1-\alpha_{\max}m_2/E_2}{C_{\max}}=2.49992 .
\label{S-escape-comparison}
\end{equation}
Thus
\begin{equation}
|b_2|
<
\frac{1-\alpha_{\max}m_2/E_2}{C_{\max}},
\label{S-escape-inequality}
\end{equation}
in agreement with the positive value of
$\delta_{\rm esc}$ in
Eq.~\eqref{S-example-compatibility-margins}.

Although fragment 2 is initially directed inward, numerical
integration reveals an inner radial turning point
\begin{equation}
r_{2,\mathrm{turn}}\simeq0.012,\qquad
\mathcal V_2(r_{2,\mathrm{turn}})=0 .
\label{S-positive-turning-point}
\end{equation}
After reversing its radial motion, the fragment crosses $r_*$ on the
outward branch. The strict no-barrier inequality in
Eq.~\eqref{S-escape-inequality} guarantees
$\mathcal V_2>0$ throughout the exterior domain
$\mathcal D_{\rm ext}=\{r\geq r_*\}$. Hence no further radial turning
point occurs, and the amplified fragment reaches the asymptotically
flat end.

The extracted Killing energy and efficiency are
\begin{equation}
\Delta E=E_2-E_0=-E_1=0.0214,\,
\eta=\frac{\Delta E}{E_0}\simeq2.14\%.
\label{S-efficiency}
\end{equation}

\end{document}